# Creating big time crystals with ultracold atoms

**Krzysztof Giergiel[1], Tien Tran[2], Ali Zaheer[2], Arpana Singh[2], Andrei Sidorov[2], Krzysztof Sacha[1] and Peter Hannaford[2]**

[1] Instytut Fizyki Teoretycznej, Uniwersytet Jagielloński, ulica Profesora Stanisława Lojasiewicza 11, PL-30-348 Krakow, Poland

[2] Optical Sciences Centre, Swinburne University of Technology, Hawthorn, Victoria 3122, Australia

E-mail: phannaford@swin.edu.au; krzysztof.sacha@uj.edu.pl

**Abstract.** We investigate the size of discrete time crystals $s$ (ratio of response period to driving period) that can be created for a Bose-Einstein condensate (BEC) bouncing resonantly on an oscillating mirror. We find that time crystals can be created with sizes in the range $s \approx 20 - 100$ and that such big time crystals are easier to realize experimentally than a period-doubling ($s = 2$) time crystal because they require either a larger drop height or a smaller number of bounces on the mirror. We also investigate the effects of having a realistic soft Gaussian potential mirror for the bouncing BEC, such as that produced by a repulsive light-sheet, which is found to make the experiment easier to implement than a hard-wall potential mirror. Finally, we discuss the choice of atomic system for creating time crystals based on a bouncing BEC and present an experimental protocol for realizing big time crystals. Such big time crystals provide a flexible platform for investigating a broad range of non-trivial condensed matter phenomena in the time domain.

## 1. Introduction

In 2012 Frank Wilczek proposed that a quantum many-body system in the lowest state could spontaneously break time-translation symmetry to form a time crystal, in analogy with the formation of a crystal in space [1]. Although such time crystals cannot exist in the lowest state of a quantum system with two-body interactions [2, 3] (see [4] for possible time crystals with long-range multi-particle interactions), it was later demonstrated that a periodically driven many-body quantum system can spontaneously break discrete time-translation symmetry to form a discrete time crystal which evolves with a period two-times ($s = 2$) longer than the driving period [5]. Such a time crystal is predicted to be robust against external perturbations and to persist perpetually in the limit of a large number of particles [6, 7]. Similar ideas of discrete time crystals were later proposed for periodically driven *spin* systems [8-10], which in the case of a spin-1/2 system evolve with a period twice as long as the driving period. Experimental evidence of discrete time crystals has since been reported for a range of spin systems, including a spin chain of ions [11], nitrogen-vacancy spin impurities in diamond [12] and nuclear spins in organic molecules [13] and ordered ADP crystals [14, 15]. In addition, space-time crystals − with periodicity in both space and time − have been reported for a superfluid Bose-Einstein condensate (BEC) of atoms [16, 17]. Experiments demonstrating spontaneous emergence of periodic evolution that does not require periodic driving have also been performed in magnon BECs consisting of bosonic quasi-particles [18, 19]. In Refs. [20, 21] it has been shown that periodically driven systems can also reveal crystalline structures in phase space. A number of comprehensive reviews on time crystals have recently been published [22-25].

In a recent paper [6] we presented mean-field calculations for a BEC of attractively interacting atoms bouncing resonantly on an oscillating mirror that exhibited dramatic breaking of time-translation symmetry to form a discrete time crystal. These time crystals can evolve with a period more than an order of magnitude longer ($s >> 10$) than the driving period, thereby creating a large number of available 'lattice sites' in the time domain. Such a system provides a flexible platform for investigating a broad range of nontrivial condensed matter phenomena in the time domain [6, 26-31]. Other time crystal systems with large values of $s$ have also recently been proposed [32-34].

In this paper we investigate the range of sizes ($s$-values) of discrete time crystals that can be created for a BEC of ultracold atoms bouncing resonantly on an oscillating mirror. We find that time crystals can be created in the range $s \approx 20 - 100$ and, furthermore, that such big time crystals are much easier to realize experimentally than a period-doubling time crystal ($s = 2$) because they require either a larger drop height or a smaller number of bounces on the mirror. We also investigate the effects of having a realistic soft Gaussian potential atom mirror (rather than a theoretical hard-wall mirror) – such as that produced by a repulsive light-sheet – which allows us to operate with a much larger mirror oscillation amplitude and in turn makes the experiment easier. Finally, we discuss the choice of atomic system for creating time crystals based on a bouncing BEC and present an experimental protocol for realizing big time crystals.

## 2. Theoretical

### 2.1 Single-particle case for a hard-wall mirror

We first consider a single atom bouncing in the vertical direction $z$ on a harmonically oscillating hard-wall mirror in the presence of strong transverse harmonic confinement. Introducing gravitational units $l_0 = (\hbar^2/(m^2 g))^{1/3}$, $E_0 = mgl_0$, $t_0 = \hbar/(mgl_0)$ and assuming a one-dimensional (1D) approximation, the Hamiltonian of the system in the laboratory frame can be written

$$H = p^2/2 + \tilde{V}[z + (\lambda/\omega^2) z \cos(\omega t)] + z, \tag{1}$$

where the potential $\tilde{V}(z \leq 0) = \infty$ and $\tilde{V}(z > 0) = 0$ for the hard-wall mirror, $\omega$ and $\lambda/\omega^2$ are the frequency and the amplitude of the oscillating mirror in the laboratory frame, respectively, and $m$ and $g$ are the atom mass and gravitational acceleration. Use of gravitational units allows calculations to be performed with dimensionless parameters that are independent of the mass of the atom.

Transforming from the laboratory frame to the oscillating frame of the mirror, Eq. (1) becomes [5, 35]

$$H = p^2/2 + \tilde{V}(z) + z + \lambda z \cos(\omega t), \tag{2}$$

where λ is the amplitude of the time-periodic perturbation in the oscillating frame (hereafter referred to as the amplitude of the oscillating mirror). The case of a realistic soft Gaussian potential mirror is considered in Section 2.6.

In the classical description, transforming to action-angle variables $(I, \theta)$

$$z = \tfrac{1}{2}(3\pi I)^{2/3}\left[1 - \left(\tfrac{\theta}{\pi}\right)^2\right]; \quad p = -\left(\tfrac{3I}{\pi^2}\right)^{1/3}\theta, \tag{3}$$

where $0 \le \theta < 2\pi$, and using the secular approximation [35,36], Eq. (2) can be expressed as a single-particle time-independent Hamiltonian [6]

$$H_F \approx P^2/(2m_{\text{eff}}) + \lambda\langle z\rangle\cos(s\Theta), \tag{4}$$

where $\Theta = \theta - \omega s/t$, $P = I - I_s$, $I_s = \pi^2 s^3/(3\omega^3)$ is the action of the $s:1$ resonant orbit. The effective mass and classical average value of $z$ are given by

$$m_{\text{eff}} = -\frac{\pi^2 s^4}{\omega^4} = -\left(\tfrac{9}{\pi}\right)^{2/3} I_s^{4/3}; \quad \langle z\rangle = \frac{1}{\omega^2} = \frac{1}{s^2}\left(\tfrac{3I_s}{\pi^2}\right)^{2/3}, \tag{5}$$

where $s = \omega/\Omega$ and $\Omega$ is the bounce frequency of the unperturbed particle (i.e., for a static mirror). The height of the classical turning point (drop height) of the bouncing atom is then $h_0 = \pi^2/(2\Omega^2)$.

Equation (4) is the effective Hamiltonian of a particle in the frame moving along an $s:1$ resonant orbit and indicates that for $s \gg 1$ a single resonantly driven atom in the vicinity of a resonant trajectory behaves like an electron moving in a crystalline structure created by ions in a solid state system. That is, eigenvalues of the quantized version of the Hamiltonian (4) form energy bands (see Fig. 1) and the corresponding eigenstates are Bloch waves. These eigenvalues are actually quasi-energies of a periodically driven particle while the eigenstates are Floquet states obtained in the frame moving along the resonant orbit [5, 6, 26, 35]. In the quantum description we will apply the tight-binding approximation by restricting the analysis to the first energy band of the quantized version of the Hamiltonian (4). In the *many-body* case such an approximation is valid provided the interaction energy per particle is smaller than the energy gap $\Delta E$ between the first and second energy bands shown in Fig. 1.

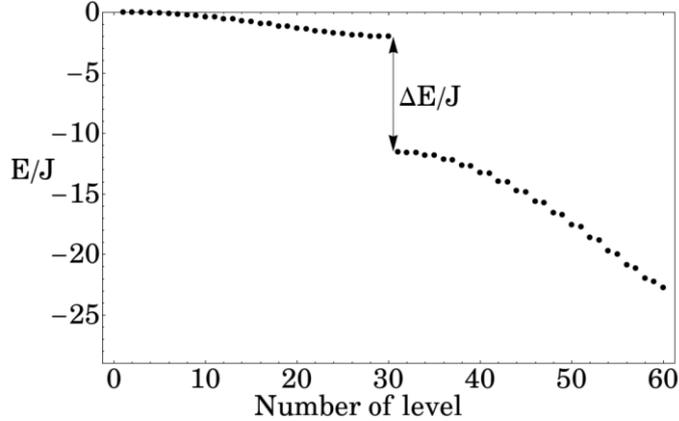

**Figure 1.** Eigenenergies $E$ for the first and second energy bands of the quantized version of the Hamiltonian (4) for $s = 30$. Because the effective mass in (4) is negative, cf. (5), the first energy band is the highest on the energy scale. The mirror oscillation frequency $\omega = 4.45$ and the oscillation amplitude $\lambda = 0.2$ are chosen so that the energy gap $\Delta E / J \approx 10$. The energies are given in units of the nearest-neighbour tunnelling amplitude $J$ corresponding to the first energy band.

### 2.2 Optimal value of the driving strength λ for a hard-wall mirror

We estimate the largest $\lambda$ for a hard-wall potential mirror that is allowed before the dynamics become chaotic, making use of the Chirikov criterion [37]. This criterion estimates the value of $\lambda$ for which two neighbouring resonance islands, $s : 1$ and $(s + 1) : 1$, described by Eq. (4), overlap. The distance between the islands (for $s \gg 1$) and the half-width of the islands are

$$I_{s+1} - I_s \approx (\pi^2/\omega^3)\, s^2; \quad \tfrac{1}{2}(\Delta I)_s = 2\sqrt{|m_{\text{eff}}\lambda\langle z\rangle|} = \sqrt{\lambda}\, 2\pi s^2/\omega^3. \tag{6}$$

The resonance islands nearly overlap when

$$\tfrac{2}{3}(I_{s+1} - I_s) \approx \tfrac{1}{2}(\Delta I)_{s+1} + \tfrac{1}{2}(\Delta I)_s, \tag{7}$$

which leads to $\lambda \approx (\pi/6)^2 \approx 0.27$. The 2/3 factor in (7) is an empirical correction that allows for the presence of higher-order resonances [35]. This means that for $s \gg 1$ the critical value of $\lambda$ is a constant independent of $s$ and $\omega$. For $\lambda = 0.2$, we find that, although there is some chaos between the neighbouring $s : 1$ and $(s + 1) : 1$ islands, the islands themselves are still not perturbed [6]. For $\lambda < 0.2$, the resonance islands become smaller and less suitable for realization of a time crystal. Indeed, quantum states that describe time crystals are located inside the resonance islands and if the islands are too small we need to choose a large value of $I_s$ in order to realize a time crystal. While this is in principle possible, the resulting evolution of ultracold atoms that demonstrates time crystal behaviour becomes very long, see Section 2.4. We conclude that for a hard-wall potential mirror $\lambda = 0.2$ is universally good for any $s : 1$ resonance for which $s \gg 1$.

### 2.3 Scaling of parameters with $s = \omega/\Omega$

The quantum secular approximation allows us to obtain the quantum version of the classical Hamiltonian (4) [6]. That is, switching to the oscillating frame by means of the unitary transformation $U = e^{i\hat{n}\omega t/s}$, the matrix elements of the time-averaged quantum Hamiltonian describing the $s:1$ resonance dynamics reads

$$\langle n'|H_F|n\rangle \approx \left(E_n - \frac{n\omega}{s}\right)\delta_{n',n} + \frac{\lambda}{2}\langle n'|z|n\rangle\left(\delta_{n',n+s} + \delta_{n',n-s}\right), \qquad (8)$$

where the $|n\rangle$ are eigenstates of the unperturbed part of the Hamiltonian (2), i.e., $H$ with $\lambda = 0$, with the corresponding eigenvalues $E_n$, and $\hat{n}|n\rangle = n|n\rangle$. The validity of the quantum secular approximation (8) can be checked for a given set of parameters by comparing results of the classical secular approach with the exact classical approach. That is, if the classical secular and exact treatments agree, the quantum secular approach is also valid [6]. Around the resonant value of the quantum number of the unperturbed particle, i.e., for $n \approx n_0 \approx I_s$, the first term on the right hand side of (8) can be approximated by

$$E_n - \frac{n\omega}{s} \approx E_{n_0} - \frac{n_0\omega}{s} + \frac{(n-n_0)^2}{2m_{\text{eff}}}. \qquad (9)$$

We know how the parameters of Eq. (4) scale with $I_s$. We now investigate how the effective mass $m_{\text{eff}}$, obtained from the quantum approach (8), and the matrix element $\langle n_0 - s/2|z|n_0 + s/2\rangle$, which provides an estimate of the classical average value $\langle z \rangle$, scale with $n_0$ and $s$. The results, presented in Fig. 2, show that for $s \leq 100$, the classical scaling is reproduced in the quantum approach if the particle quantum number $n_0 \gtrsim 100$. This allows us to use the classical analytical expressions to determine the optimal parameters for an experiment.

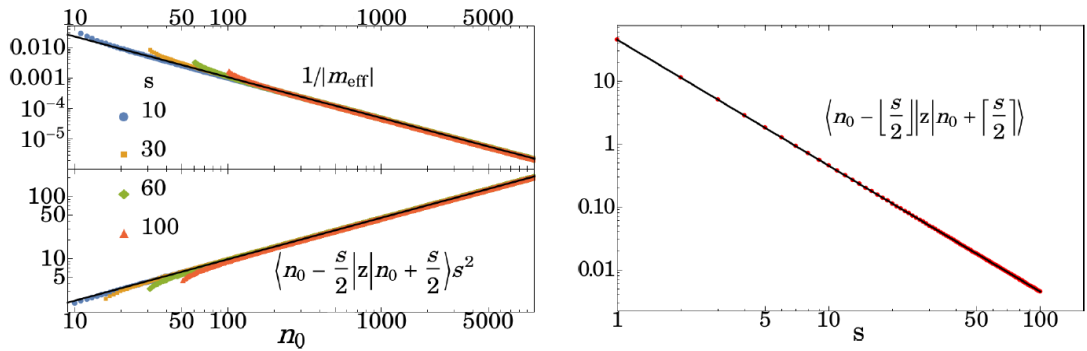

**Figure 2**. Left panel: inverse effective mass $1/|m_{\text{eff}}|$ and the matrix element $\langle n_0 - s/2|z|n_0 + s/2\rangle$, which provides an estimate of the classical average $\langle z \rangle$ in Eq. (5), versus particle quantum number $n_0$ for different $s:1$ resonances in the range $s = 10 - 100$ (as indicated in the plot) for a hard-wall potential mirror. The classical results (5) (black curves) predict $1/|m_{\text{eff}}| \propto n_0^{-4/3}$ and $\langle z \rangle \propto n_0^{2/3}$ which are observed in the quantum approach for $n_0 \gtrsim 100$ for any $10 \leq s \leq 100$. Right panel: matrix element $\langle n_0 - s/2|z|n_0 + s/2\rangle$ versus $s$ for $n_0 = 1000$. The classical result (5) (black curve), which predicts $\langle z \rangle \propto s^{-2}$ for a fixed $I_s \approx n_0$, is observed in the quantum approach (red points).

*2.4 Number of bounces and energy gap*

To demonstrate that a time crystal is created one needs to show that the ultracold atoms evolve periodically with a period $s$-times longer than the driving period $T = 2\pi/\omega$. If the interactions between atoms are too weak, the subharmonic periodic evolution will be destroyed because the atoms will start to tunnel between lattice sites of the potential in (4), or in other words between wave-packets propagating along the resonant orbit. Therefore, to demonstrate that a time crystal is created, the system needs to have evolved for at least the time period corresponding to the tunnelling time $t_{\text{tunnel}}$ of a single atom between neighbouring lattice sites. The tunnelling time (for $s >> 1$) and the bounce period $T_{\text{bounce}}$ of the atom are given by [6]

$$t_{tunnel} \approx 2.4/J; \quad T_{bounce} = 2\pi/\Omega, \tag{10}$$

where $J$ is the tunnelling amplitude of the particle between neighbouring sites of the periodic potential in (4), $\Omega = [\pi^2/(3n_0)]^{1/3}$, and $I_s$ is denoted here by the quantum number of the unperturbed particle, i.e., $I_s \approx n_0$. We require $t_{\text{tunnel}}$ to be as short as possible because then the number of bounces needed to demonstrate that a time crystal is robust against single-particle tunnelling is relatively small – each bounce off the mirror is a potential source of loss of atoms from the BEC. The number of bounces required to observe quantum tunnelling for non-interacting particles is then

$$N_b = \frac{t_{tunnel}}{T_{bounce}} \approx 2.4\Omega/(2\pi J) \text{ for } s >> 1. \tag{11}$$

Another important parameter that we wish to control is the energy gap $\Delta E$ between the first and second bands of the quantum version of the Hamiltonian (4), see Fig. 1. To realize a time crystal, sufficiently strong interactions between atoms need to be be present. However, if the interactions are too strong, the single-band description within the Bose-Hubbard model (see Section 2.5) is no longer valid because higher bands of the Hamiltonian (4) become involved. Thus, $\Delta E$ should be as large as possible but this requirement is in contradiction with the requirement of a short tunnelling time $t_{\text{tunnel}}$ and we need to find a compromise.

Dividing the classical Hamiltonian (4) by $\lambda\langle z\rangle$ and rescaling $s\Theta \rightarrow \Theta$, gives

$$H_F \approx P^2/[2m_{\text{eff}}\lambda\langle z\rangle/s^2] + \cos\Theta = P^2/\alpha + \cos\Theta, \tag{12}$$

which depends only on the single universal parameter

$$\alpha = 2|m_{\text{eff}}|\lambda\langle z\rangle/s^2 = 2\pi^2 s^2\lambda/\omega^6. \tag{13}$$

The division by $\lambda\langle z\rangle$ also means that in order to analyse the scaling properties of the system the quasi-energy gap $\Delta E$ and the tunnelling amplitude $J$ should be expressed in units of $\epsilon = \lambda\langle z\rangle = \lambda/\omega^2$.

Figure 3 (top panel) presents calculations of the energy gap $\Delta E/J$, obtained from the quantum secular Hamiltonian (8), cf. Fig. 1, versus the universal parameter $\alpha$ for a range of values of $s$ and $n_0$ for a hard-wall potential mirror. We see that for different values of $s$ and $n_0$ all the quantum data lie on a single universal curve. Similarly, in Fig. 3 (bottom panel), the tunnelling amplitude $J$, in units of $\epsilon = \lambda\langle z\rangle$, obtained from diagonalization of the effective Hamiltonian (4) and from the fully quantum secular approach (8), versus $\alpha$ for different values of $s$ and $n_0$ all lie on the same universal curve which we denote by $f(\alpha) = J(\alpha)/\epsilon$. These results demonstrate that the scaling deduced from our analysis is valid and can be used for further purposes.

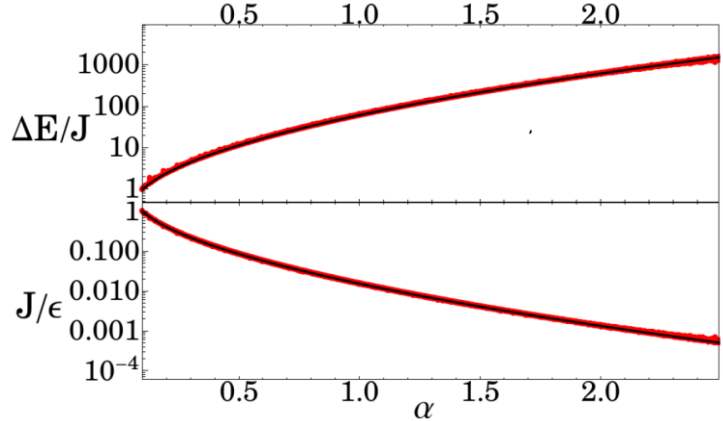

**Figure 3.** Top panel: energy gap $\Delta E$ between the first and second bands (cf. Fig. 1), in units of the tunnelling amplitude $J$, versus the universal parameter $\alpha = 2|m_{\mathrm{eff}}|\lambda\langle z\rangle/s^2$. Bottom panel: tunnelling amplitude $J$, in units of $\epsilon = \lambda\langle z\rangle$, versus the universal parameter $\alpha$. The black curves show the results of the diagonalization of the quantized version of the classical effective Hamiltonian (12). The red points are related to the fully quantum secular approach, Eq. (8), for $10 \leq s \leq 100$, $0.007s^3 \leq n_0 \leq 0.06s^3$ and $\lambda$ chosen so that $\alpha$ is in the range $0 - 2.5$.

With the help of $f(\alpha) = J(\alpha)/\epsilon$, we can now express the number of bounces as

$$N_b = \frac{t_{tunnel}}{T_{bounce}} = 2.4/[\sqrt{2\lambda\alpha}\, f(\alpha)], \tag{14}$$

which indicates that if we choose $\lambda$ (e.g., the optimal value $\lambda = 0.2$ for a hard-wall potential mirror) and a value for the energy gap $\Delta E(\alpha)/J(\alpha)$ (which determines $\alpha$) the number of bounces $N_b$ is constant and independent of the values of $s$ that we choose. In other words, for fixed $\lambda$ and $\alpha$ we can choose any $s : 1$ resonance (then from $\alpha = $ constant we obtain $n_0$ and consequently $\omega$) and the number of bounces is always the same. This is illustrated in the left panel of Fig. 4 where for different values of $\lambda$ we always choose $\alpha$ so that the gap $\Delta E/J$ between the first and second energy bands of the quantum version of the Hamiltonian (4) is about 10 (which for $\lambda = 0.2$ corresponds to $\alpha = 0.456$).

The key results of the analysis are presented in the right panel of Fig. 4 where the number of bounces $N_b$ needed for tunnelling of non-interacting atoms between lattice sites of the potential in (4) versus the band gap $\Delta E/J$ are presented. The smaller the band gap we can afford, the smaller the optimum number of bounces.

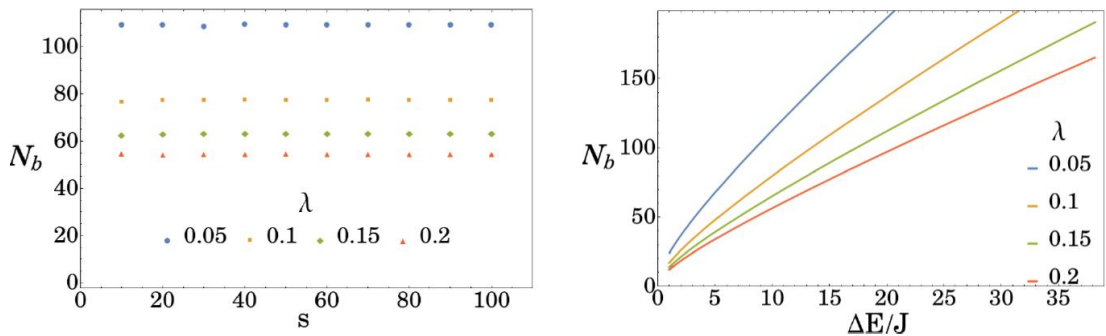

**Figure 4.** Left panel: number of bounces $N_b$ needed to observe tunnelling of non-interacting atoms between lattice sites of the potential in (4). These are optimal values, i.e., the smallest possible values with the restriction that the energy gap $\Delta E/J \approx 10$. The results are presented versus $s = \omega/\Omega$ for different amplitudes $\lambda$ of the oscillations of the hard-wall potential mirror, where $n_0$ is always chosen so that $\Delta E/J = 10$. Right panel: number of bounces $N_b$ for different oscillation amplitudes $\lambda$ versus energy gap $\Delta E/J$.

## 2.5 Many-body case

We now switch to the case of a BEC of interacting bosonic atoms bouncing on an oscillating mirror in the presence of strong transverse harmonic confinement $\omega_\perp$ and assume the 1D approximation. Note that we focus on a spatially finite 1D system where for sufficiently weak interactions, the BEC is not destroyed by long-wavelength quantum fluctuations. We restrict the analysis to the Hilbert subspace corresponding to the first energy band of the quantum version of the Hamiltonian (4). This is the resonant subspace where atoms occupy $s$ localized wavepackets $w_i(z,t)$ evolving along the $s : 1$ resonant orbit with period $sT$, where $T$ is the driving period [6]. These wave-packets are the time-periodic version of the Wannier states in solid state physics. Within the mean-field approach and restricting to the $s$-dimensional resonant Hilbert subspace we can expand solutions of the Gross-Pitaevskii equation in terms of localized Wannier-like states $\psi(z,t) = \sum_{i=1}^{s} a_i w_i(z,t)$ and obtain the energy functional (actually the quasi-energy functional) in the form [6]

$$E \approx -\frac{1}{2}\sum_{i,j=1}^{s} J_{ij}\left(a_i^* a_j + c.c.\right) + \frac{1}{2}\sum_{i,j=1}^{s} U_{ij}|a_i|^2|a_j|^2, \qquad (15)$$

with

$$J_{j'j} = -\frac{2}{sT}\int_0^{sT} dt \int_0^\infty dz\, w_{j'}^*\left[\frac{p^2}{2} + z + \lambda z\cos\omega t - i\partial_t\right]w_j,$$

$$U_{ii} = \frac{g_{1D}N}{sT}\int_0^{sT} dt \int_0^\infty dz\, |w_i|^4,$$

$$U_{ij} = \frac{2g_{1D}N}{sT}\int_0^{sT} dt \int_0^\infty dz\, |w_i|^2|w_j|^2, \quad for \quad i \neq j,$$

where $N$ is the total number of atoms, $g_{1D} = 2\omega_\perp a_s$ describes the contact interaction between the atoms, $a_s$ is the $s$-wave scattering length (in gravitational units) and the $U_{ii}$ and $U_{ij}$ describe the on-site and long-range interaction energies per particle in the Bose-Hubbard model (15). The leading tunnelling amplitudes correspond to the nearest-neighbour hopping $J = |J_{i,i+1}|$ and only this hopping can be kept in the model when we want to describe time crystal dynamics. Such a tight-binding approximation is valid provided the interaction energy per particle is much smaller than the energy gap $\Delta E$ between the first and second bands of the effective Hamiltonian (4), see Fig. 1.

In the left panel of Fig. 5 the lowest energy solutions of the mean-field energy (15) for $s = 30$, $\lambda = 0.2$, $\omega = 4.45$ and different values of the interaction strength $g_{1D}N$ are presented. We emphasize that these solutions are related to the lowest quasi-energy of the driven system within the resonant Hilbert subspace; they do not correspond to the ground state which does not exist in a periodically driven system. For $g_{1D}N > -0.0024$, the lowest energy solution is a uniform superposition of all $s$ Wannier-like wave-packets, $\psi = \sum_{i=1}^{s} w_i/\sqrt{s}$. The wavefunction $\psi$ evolves with the driving period $T$ and no spontaneous breaking of discrete time-translation symmetry of the many-body Hamiltonian takes place. However, if the attractive interactions are sufficiently strong, it becomes energetically favourable for the atoms to localize, and the lowest energy mean-field solution reveals spontanoues breaking of time-translation symmetry; i.e., $\psi$ evolves with a period $s$-times longer than $T$. The stronger the interactions, the better the localization of the atoms, and for $g_{1D}N \le -0.07$, we obtain $\psi \approx w_i$ with the squared-overlap $|\langle\psi|w_i\rangle|^2 > 0.9$. The corresponding interaction energy per particle $|U_{ii}|/(2J) \approx 0.8$ (where $J = |J_{i,i+1}|$) is smaller than the energy gap $\Delta E/J = 10$ between the first and second bands of the quantum version of the Hamiltonian (4), and consequently the Bose-Hubbard model (15) is valid.

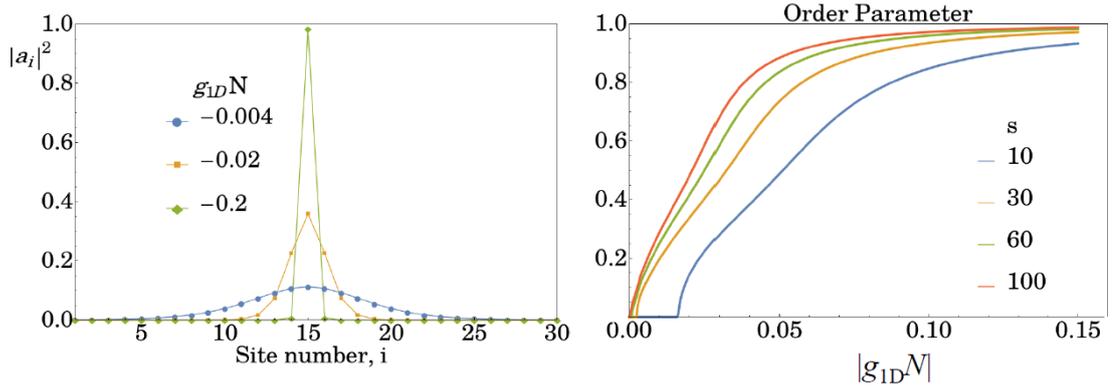

Figure 5. Left panel: occupation probabilities of localized Wannier wave-packets $w_i$ in the lowest energy solutions of the mean-field Gross-Pitaevskii equation corresponding to the energy functional (15) for the interaction strengths $g_{1D}N$ indicated in the plot and $s = 30$, $\lambda = 0.2$, $\omega = 4.45$, for a hard-wall potential mirror. For $g_{1D}N \le -0.0024$, the time-translation symmetry is broken. For $g_{1D}N = -0.2$ (green diamonds), the atoms are localized in a single Wannier wave-packet. Right panel: order parameter (16) versus $|g_{1D}N|$ for different $s : 1$ resonances, $\lambda = 0.2$ and $\omega$ (or equivalently $n_0$) chosen so that $\Delta E/J = 10$.

In the right panel of Fig. 5 the order parameter

$$O = \frac{\text{Max}_i\left(|\langle\psi|w_i\rangle|^2\right)-1/s}{1-1/s},$$ (16)

which characterizes the overlap of the lowest energy solution $\psi$ with a single Wannier-like wave-packet $w_i$ versus the interaction strength $g_{1D}N$ is presented for different $s$. The critical values of the interaction strength can be identified in the plots. The figure also shows how strong the interactions need to be in order to be dealing with the lowest energy solution that practically reduces to a single Wannier-like wave-packet $w_i$. This is important information because time crystals in which $\psi(z, t) \approx w_i(z, t)$ can be easily prepared in an experiment, see Section 4. The case $s = 40$ has been extensively analyzed in [6], where the numerical integration of the full Gross-Pitaevskii equation confirmed the results based on the Bose-Hubbard model.

For different $s$ : 1 resonances, the critical interaction strength $g_{1D}N$ differs because the tunnelling amplitudes $J_{ij}$ are slightly different and also because the same $g_{1D}N$ does not necessarily mean the same interaction coefficients $U_{ij}$ in (15). The coefficients $U_{ij}$ depend on the longitudinal width of the atom cloud which, for the optimal values of the parameters, varies from $\sigma_z = 1.3 - 2.7$ for $s = 10 - 100$ (Table 2). Thus, a slightly different interaction parameter $g_{1D}N$ and transverse confinement frequency $\omega_\perp$ is required for different $s$.

## 2.6 Case of a soft Gaussian potential mirror

The calculations in previous sections were based on a simple hard-wall mirror potential. We now consider the case of a realistic soft Gaussian mirror potential, such as that produced by a repulsive light-sheet,

$$V(z) = V_0 \, exp(-z^2/2\sigma_0^2),$$ (17)

where $V_0$ and $\sigma_0$ are the height and width of the Gaussian mirror potential.

For a hard-wall potential mirror, the trajectories of the bouncing atoms reverse their direction abruptly at the reflection point (i.e., $p \to -p$), so that the Fourier transform of the unperturbed periodic trajectories $z(t) = \sum_k z_k e^{ik\Omega t}$ (where $\Omega$ is the frequency of the bouncing atom in the absence of mirror oscillations, see (10)) results in amplitudes of the harmonics that decrease with $k$ like $z_k \sim 1/k^2$. In the case of a soft Gaussian potential mirror, the trajectories of the atoms are smoothly reflected, so that the harmonics decrease much faster with $k$, see left panel of Fig. 6. Because for an $s$ : 1 resonance the amplitude of the potential in (4) is determined by the amplitude of the $s$-th harmonic, i.e., $\langle z \rangle \approx z_s$, the Gaussian potential mirror needs to oscillate with an amplitude $\lambda$ much larger than the oscillation amplitude of a hard-wall potential mirror in order to have the same effect on the bouncing atoms.

When a particle bounces off a soft Gaussian potential mirror some harmonics of the classical unperturbed orbits are not created at all. The question of which harmonics disappear for a given $V = V_0/E_{\text{particle}} \geq 1$ (where $E_{\text{particle}}$ is the particle's energy) is a complex problem. In the left panel of Fig. 6 we see dramatic drops of certain Fourier components which have a uniform spacing increasing from $\Delta k = 16 - 33$ for $V \approx 1.6 - 5$. We focus here on the $30$ : 1 resonance. The right panel of Fig. 6 shows the amplitude $z_{30}$ of the $30^{\text{th}}$ harmonic of an unperturbed periodic orbit as a function of $V = V_0/E_{\text{particle}}$. When $V$ approaches one, i.e., $V_0 = E_{\text{particle}}$, from above there is a series of zeros of $z_{30}$. However, for $V \gtrsim 10$, there is no dramatic drop of $|z_{30}|$ and an experiment demonstrating a discrete time crystal can be carried out in this regime.

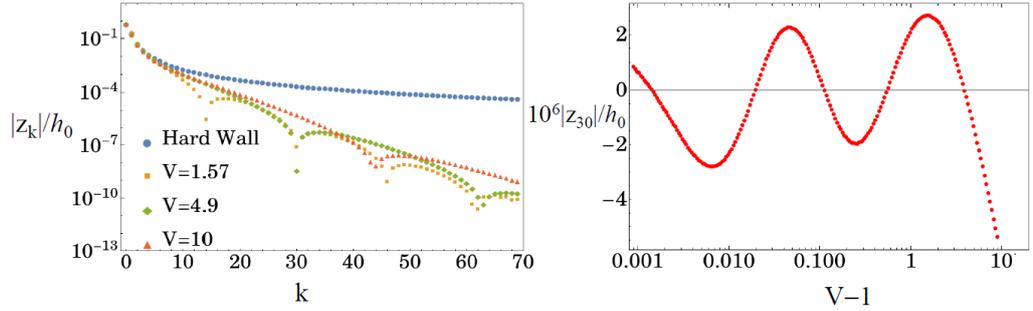

**Figure 6.** Left panel: modulus of Fourier components $z_k$ of the unperturbed classical orbits, $z(t) = \sum_k z_k e^{ik\Omega t}$, in units of the drop height $h_0$, for a particle bouncing on a static hard-wall potential mirror (blue dotted line) and a static Gaussian potential mirror with different mirror heights $V = V_0/E_{\text{particle}}$ in units of the particle energy. In the hard-wall mirror case, the Fourier components decrease monotonically like $z_k \sim 1/k^2$. In the case of the soft Gaussian mirror, the decrease of $z_k$ is much faster and some $z_k$ components drop abruptly. Right panel: Fourier component $z_{30}$ of the trajectory of an atom bouncing on a static Gaussian mirror versus $V - 1$. A positive $V - 1$ signifies that the barrier of the Gaussian mirror is larger than the particle energy. For some values of $V$, the $z_{30}$ component disappears and there will be no $30:1$ resonance islands when we drive an atom with frequency $\omega = 30\Omega$.

For a given choice of $V_0$ and $\sigma_0$, our primary restriction is to obtain an energy gap of $\Delta E/J \approx 10$. First, we choose a mirror oscillation frequency $\omega$ and by applying the quantum secular approximation (8) we determine the optimal oscillation amplitude $\lambda$ that leads to $\Delta E/J \approx 10$. We then need to check if the secular approximation is still valid for these parameters by examining the classical phase-space pictures of the action $I$ versus angle $\Theta$ to see if the dynamics is still regular or if it is already chaotic.

We assume here a soft Gaussian mirror potential which corresponds approximately to the repulsive light-sheet mirror used in [38] for the reflection of a $^{87}$Rb BEC dropped from heights up to 300 µm. The mirror is formed by a ≤ 3W, 532 nm laser beam with a waist $\sigma_0 l_0 = 10$ µm and horizontal extension 200 µm, which for $^{39}$K atoms corresponds to $\sigma_0 = 15.5$ and $V_{\text{max}} \approx 4.6 \times 10^3$ in gravitational units. The hardness of the mirror can be varied by varying the beam waist $\sigma_0$. For a given Gaussian width $\sigma_0 = 15.5$, different mirror potential heights $V_0/V_{\text{max}}$, and oscillation frequencies around the optimal hard-wall mirror value $\omega = 4.45$, we have used the quantum secular approach (8) to determine the amplitude of the mirror oscillations $\lambda$ required to obtain $\Delta E/J \approx 10$. The results are shown in the left panel of Fig. 7. Now that we have predictions for all parameters, the corresponding classical phase-space pictures have been obtained, examples of which are shown in Fig. 8 for different mirror heights $V_0/V_{\text{max}}$. For $\omega = 4.45$, the phase space around the $30:1$ resonance islands is regular in all cases except $V_0/V_{max} \approx 0.2$, for which the $\lambda$ needed to obtain $\Delta E/J \approx 10$ is extremely high ($> 100$) and the classical motion is no longer regular, and consequently the $30 : 1$ resonance for $V_0/V_{max} \approx 0.2$ is not suitable for realization of a time crystal. This is because for $V_0/V_{max} \approx 0.2$ (and for a particle energy that fulfills $V = V_0/E_{particle} \approx 5$), the $30^{th}$ harmonic of the resonant periodic orbit disappears (Fig. 6). For $V_0/V_{max} > 0.2$, the drop height $h_0$ does not change too much as a function of $V_0/V_{max}$ (Fig. 7, right panel) and the optimal number of bounces required for tunnelling to neighbouring wave-packets and the optimal tunnelling amplitude remain nearly constant at $N_b \approx 54$ and $J \approx 0.0010$ for $\omega = 4.45$. All of these parameters are very close to the optimal values for the hard-wall mirror.

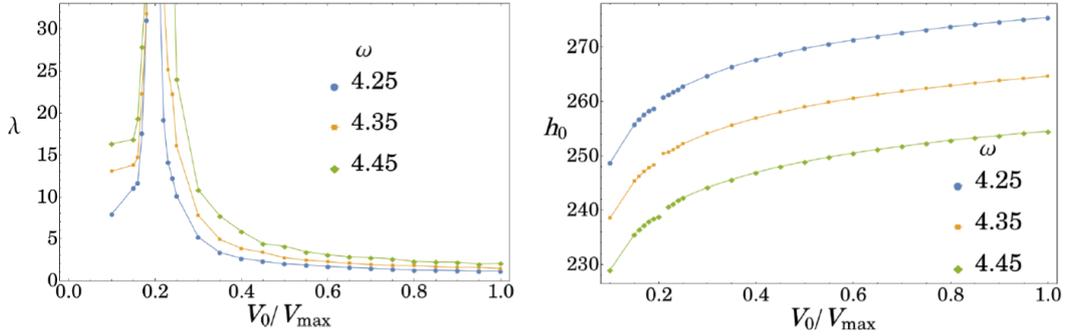

**Figure 7.** Left panel: optimal values of the oscillation amplitude $\lambda$ needed to obtain $\Delta E/J = 10$ versus height of the Gaussian mirror potential $V_0/V_{max}$ for different mirror oscillation frequencies $\omega$ (as indicated in the plots) and $\sigma_0 = 15.5$, $V_{max} \approx 4.6 \times 10^3$. Right panel: corresponding drop heights $h_0$ versus $V_0/V_{max}$. The results are obtained within the quantum secular Hamiltonian (8) and need to be validated by checking the classical phase-space pictures like in Fig. 8.

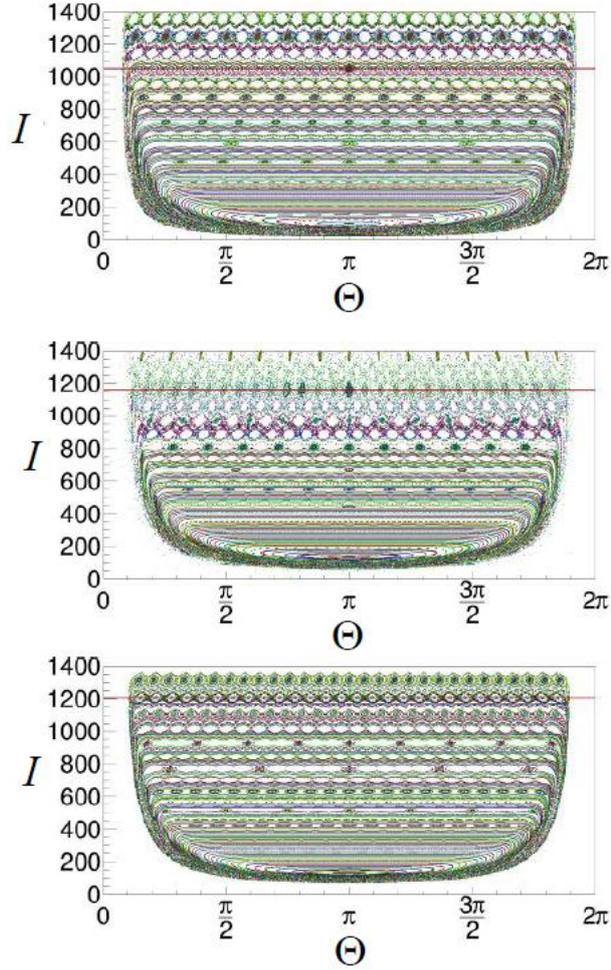

**Figure 8.** Classical phase-space pictures of the action $I$ versus angle $\Theta$ for the reflection of atoms from a soft Gaussian mirror potential with different mirror potential heights $V_0/V_{max} = 0.1$, $0.2$, $0.8$ (top to bottom), where $V_{max} \approx 4.6 \times 10^3$ and $\sigma_0 = 15.5$, $\omega = 4.45$. The mirror oscillation amplitude $\lambda$ is chosen so that the quantum secular approximation (8) predicts a band gap $\Delta E/J = 10$. The phase space around the $s = 30$ resonance islands, which are located at $I \approx 1000 - 1200$ (red horizontal lines), is regular in all cases except $V_0 \approx 0.2V_{max}$, for which the $\lambda$ needed to obtain $\Delta E/J = 10$ is extremely high (cf. Fig. 7, left panel) and the classical motion is no longer regular. Not all islands are visible because some are hidden in the lower, narrow and elongated part of the phase space. For $\Theta$ close to 0 and $2\pi$, i.e., close to the Gaussian mirror, the action $I$ drops to zero because we have used action-angle variables suitable for the hard-wall mirror (which are given analytically) rather than for the Gaussian mirror.

For $\omega > 4.45$ and $\lambda$ corresponding to $\Delta E/J = 10$, the classical motion becomes more chaotic than for $\omega = 4.45$ and the secular approach is not fully valid, while for $\omega < 4.45$ the number of bounces $N_b$ required for tunnelling to neighbouring wave-packets increases. Thus, the optimal value of $\omega$ for the soft Gaussian potential mirror is close to the optimal value $\omega = 4.45$ for the hard-wall potential mirror for $s = 30$.

The relatively large optimal values of $\lambda$ in the case of the soft Gaussian potential mirror, e.g., $\lambda = 2.3$ (corresponding to $\lambda l_0/\omega^2 = 74$ nm for $^{39}$K in the laboratory frame) at $V_0/V_{max} = 0.8$, $\omega = 4.45$, required to obtain a band gap $\Delta E/J = 10$ are more readily accessible experimentally than the optimal hard-wall mirror value $\lambda = 0.2$ (6.5 nm for $^{39}$K). If still larger mirror oscillation amplitudes $\lambda$ are required in the laboratory, one could operate closer to the peak of the lambda versus mirror potential height curve in Fig. 7, left panel, provided the corresponding classical resonance islands are not destroyed by chaos.

### 2.7 Constraints on the maximum number of atoms

To operate in the quasi-1D regime, the transverse standard deviation $\sigma_\perp$ of the atomic density needs to be less than or equal to the standard deviation $\sigma_z$ along the longitudinal direction and the interaction energy per particle should not exceed the excitation energy in the transverse directions ($\sigma_\perp$ and $\sigma_z$ are defined as the Gaussian fits to the density of the atomic cloud). The transverse width is determined by the frequency $\omega_\perp$ of the harmonic potential in the transverse directions, i.e., $\sigma_\perp = \sqrt{\hbar/(2m\omega_\perp)}$. These requirements imply that we need $\sigma_z \geq \sigma_\perp$ and $\sigma_z > |a_s|N$, where $a_s$ is the atomic $s$-wave scattering length. When these criteria are fulfilled, a BEC of ultracold atoms is also stable against 'bosenova' collapse that can occur for attractive interactions in three-dimensional space. For atoms bouncing resonantly on an oscillating mirror, the longitudinal width is smallest at the classical turning point and corresponds to $\sigma_z = 1.81$ for $s = 30$. For $^{39}$K atoms, $s = 30$, $\omega = 4.45$ and a Gaussian mirror with $V_0/V_{max} = 0.8$, $\sigma_0 l_0 = 10$ μm, the minimum longitudinal width is $\sigma_z l_0 = 1.17$ μm; for example, for $a_s l_0 = -1.7a_0$, the maximum number of atoms is $N_{max} \approx 10,000$.

To minimize losses due to three-body recombination, the mean-square atom density $\langle n^2 \rangle$ at the classical turning point needs to be less than $1/(K_3\tau_{BEC})$, where $K_3$ is the three-body recombination constant and $\tau_{BEC}$ is the lifetime of the BEC. For $^{39}$K $|1,+1\rangle$ atoms and taking $K_3 = 1.3$ (5) $\times 10^{-29}$ cm$^6$s$^{-1}$ near the zero-crossing point [38] and $\tau_{BEC} \approx 1$ s, we obtain $\langle n^2 \rangle_{max} \approx 10^{29}$ cm$^{-6}$; for example, for $\sigma_z l_0 = \sigma_\perp l_0 = 1.17$ μm at the classical turning point, the maximum number of atoms is $N_{max} \approx 16,000$.

## 3. Choice of Atomic System

In our earlier paper [6], we focussed on the $^{85}$Rb $|F=2, m_F=-2\rangle$ system, which has a broad Feshbach resonance at 155 G with a zero crossing point at 166 G. In Table 1, we compare the Feshbach resonance parameters and time crystal parameters for a hard-wall potential mirror and $s = 30$ for four bosonic alkali systems that have broad Feshbach resonances: $^{85}$Rb $|F = 2, m_F = -2\rangle$, $^{39}$K $|1, +1\rangle$, $^{39}$K $|1, -1\rangle$ and $^7$Li $|1, 0\rangle$.

Table 1

Feshbach resonance and time crystal parameters for $^{87}$Rb, $^{39}$K and $^7$Li atoms for a hard-wall potential mirror and $s = 30$.

| | $^{85}$Rb | $^{39}$K | | $^7$Li |
|---|---|---|---|---|
| **Feshbach Resonances** | $\lvert 2, -2 \rangle$ | $\lvert 1, +1 \rangle$ | $\lvert 1, -1 \rangle$ | $\lvert 1, 0 \rangle$ |
| Centre magnetic field, $B_\infty$ [G] | 155.04 [39] | 402.5 [41] | 560.7 [43] | 736.8 [44] |
| Width of resonance, $\Delta$ [G] | 10.7 | −52 | −56 | −192.3 |
| Zero crossing field, $B_0$ [G] | 165.7 | 350.5 | 504.7 | 849.9 |
| Background scattering length, $a_{bg}$ [$a_0$] | −443 | −29 | −29 | −25 |
| Background to width ratio $\lvert a_{bg}/\Delta \rvert$ [$a_0$/G] | −41.4 | 0.56 | 0.52 | 0.130 |
| $\delta B$ to give $\delta a_0 = \pm 0.1 a_0$ [mG] | 2.4 | 179 | 193 | 764 |
| 3-body recombination at $B_0$: $K_3$ [cm$^6$s$^{-1}$] | 8×10$^{-28}$[40] | 1.3(5)×10$^{-29}$[42] | 1.5(6)×10$^{-29}$[43] | 1×10$^{-27}$ [45] |
| **Gravitational units** | | | | |
| Length, $l_0$ [μm] | 0.385 | 0.647 | | 2.034 |
| Time, $t_0$ [ms] | 0.198 | 0.256 | | 0.455 |
| Energy, $E_0/k_B$ [nK] | 38.6 | 29.7 | | 16.8 |
| **Time crystal parameters ($s$=30)** | | | | |
| Mirror frequency, $\omega/(2\pi t_0)$ [kHz] | 3.6 | 2.8 | | 1.56 |
| Hard-wall mirror amplitude, $\lambda l_0/\omega^2$ [nm] | 3.9 | 6.5 | | 20.5 |
| Atom bounce period, $T_{bounce} t_0$ [ms] | 8.4 | 10.9 | | 19.3 |
| Drop height, $h_0 l_0$ [μm] | 86 | 145 | | 460 |
| Tunnel amplitude, $J/t_0$ [s$^{-1}$] | 5.3 | 4.1 | | 2.3 |
| Tunnel time, $t_{tunnel} t_0$ [s] | 0.45 | 0.59 | | 1.04 |
| Number of bounces during $t_{tunnel}$, $N_b$ | 54 | 54 | | 54 |
| Longitudinal width, $\sigma_z l_0$ [μm] | 0.70 | 1.17 | | 3.7 |
| $v_{rms}$ at 100 nK [mm/s] | 5.4 | 8.0 | | 19 |

In a time crystal experiment, the $s$-wave scattering length $a_s$ needs to be adjusted to zero for the non-time crystal phase and to small negative (attractive) values (e.g., $a_s l_0 = -1.6 a_0$) for the time crystal phase. The sensitivity of $a_s$ to magnetic fields in a Feshbach resonance is determined by the ratio of the background scattering length to the width of the resonance $a_{bg}/\Delta$. For $^7$Li and $^{39}$K, $a_{bg}/\Delta$ is 320 and 75 times smaller, respectively, than for $^{85}$Rb and therefore much less sensitive to stray magnetic fields. $^{39}$K also has the flexibility of having two broad Feshbach resonances, one involving a high-field seeking $\lvert 1, +1 \rangle$ state at $B_\infty = 402.5$ G and the other a low-field seeking $\lvert 1, -1 \rangle$ state at $B_\infty = 560.7$ G, which may be useful if the atoms need to be trapped in a magnetic trap prior to producing a BEC in an optical dipole trap.

The gravitational unit of length $l_0$ (which scales as $m^{-2/3}$) is 5.3 and 1.7 times larger for $^7$Li and $^{39}$K atoms than for $^{85}$Rb, and hence the mirror oscillation amplitude and drop height are larger by these factors. A larger mirror oscillation amplitude is more accessible in an experiment while a larger drop height allows the atom density to be probed with higher spatial resolution during a bounce cycle. On the other hand, the gravitational unit of time $t_0$ (which scales as $m^{-1/3}$) is 2.3 and 1.3 times larger for $^7$Li and $^{39}$K atoms than for $^{85}$Rb, and hence the bounce period and the time for tunnelling to neighbouring wave-packets are longer by these amounts, which makes the experiment longer compared to the lifetime of the bouncing BEC. In addition, the rms velocity (which scales as $m^{-1/2}$) for a thermal cloud is 3.5 and 1.5 larger for $^7$Li and $^{39}$K atoms than for $^{85}$Rb, which means an atom cloud needs to be 12 and 2.2 times colder than for $^{85}$Rb to have the same velocity spread.

Potassium-39 $|1, +1\rangle$ and $|1, -1\rangle$ atoms have much smaller three-body collision loss rates than $^{85}$Rb $|2, -2\rangle$ and $^7$Li $|1, 0\rangle$ atoms (Table 1), which allows higher atom densities to be used. $^{39}$K also has certain technical advantages compared with $^7$Li: high-power tapered amplifier laser systems at 767 nm (D2) and 770 nm (D1) are readily available commercially, $^{39}$K does not require a Zeeman slower, and it is easier to access the zero crossing point ($B_0 = 350$ G) than for a $^7$Li resonance ($B_0 = 545$ G) and to quickly switch or ramp the Feshbach magnetic field.

From the above considerations, we focus here on $^{39}$K as an optimal atomic system. Table 2 summarizes the parameters for $^{39}$K atoms for a hard-wall potential mirror for the range $s = 10 - 100$ and $\lambda = 0.2$, $\Delta E/J = 10$, $N_b = 54$. Ideally, we require a large mirror oscillation amplitude ($\lambda l_0/\omega^2 >> 10$ nm) which is more readily accessible in an experiment, a large drop height ($h_0 l_0 \geq 100$ μm) to allow high spatial resolution probing of the atom density during a bounce cycle, and a short tunnelling time ($< 1$ s) to allow the experiment to be performed in times shorter than the lifetime of a bouncing BEC. The results of Table 2 suggest that concerning the values of the drop height and tunnelling time, it should be feasible to create time crystals with sizes in the range $s \approx 20 - 100$.

For $s = 10$ the drop height becomes significantly smaller than 100 μm if we want to have an energy gap $\Delta E/J \approx 10$ and a number of bounces not greater than $N_b = 54$. In the extreme case of $s = 2$, even if we allow for a drop height as small as 10 μm but keep $\Delta E/J \approx 10$ (i.e., we choose $\omega = 1.1$ and $\lambda = 0.0145$), we obtain the number of bounces required for tunnelling of non-interacting atoms $N_b = 235$. Clearly, time crystals of small size are much more demanding experimentally than the big time crystals.

For the soft Gaussian potential mirror and $s = 30$, $\omega = 4.45$, we obtain $N_b \approx 54$, $h_0 \approx 253$ and thus similar parameters to the corresponding hard-wall potential mirror case. However, the mirror oscillation amplitude for the soft Gaussian potential mirror is much more accessible experimentally, i.e., ($\sim 75$ nm for $^{39}$K and a Gaussian width $\sigma_0 l_0 = 10$ μm and height $V_0/V_{max} = 0.8$, where $V_{max} \approx 4.6 \times 10^3$. Still larger mirror oscillation amplitudes could be accessed, if required, by choosing a mirror potential height $V_0/V_{max}$ closer to the peak in Fig. 7 (left panel), provided the corresponding classical resonance islands remain stable.

Table 2

Calculated parameters for $^{39}$K atom for a hard-wall potential mirror for different values of $s$ and $\lambda = 0.2$, $\Delta E/J = 10$, $N_b = 54$.

| $s$ | $\omega$ | $\omega/(2\pi t_0)$ [kHz] | $\lambda l_0/\omega^2$ [nm] | $h_0 l_0$ [μm] | $\sigma_z l_0$ [μm] | $n_0$ | $J/t_0$ [s$^{-1}$] | $t_{tunnel} t_0$ [s] |
|---|---|---|---|---|---|---|---|---|
| 10 | 3.08 | 1.92 | 13.6 | 33.5 | 0.83 | 112 | 8.5 | 0.28 |
| 20 | 3.89 | 2.42 | 8.56 | 84.5 | 1.02 | 448 | 5.3 | 0.45 |
| 30 | 4.45 | 2.77 | 6.53 | 145 | 1.17 | 1007 | 4.1 | 0.59 |
| 40 | 4.90 | 3.05 | 5.38 | 213 | 1.29 | 1790 | 3.4 | 0.71 |
| 50 | 5.28 | 3.28 | 4.64 | 281 | 1.38 | 2797 | 2.9 | 0.83 |
| 60 | 5.61 | 3.49 | 4.11 | 365 | 1.47 | 4027 | 2.6 | 0.93 |
| 70 | 5.90 | 3.67 | 3.71 | 449 | 1.55 | 5481 | 2.3 | 1.03 |
| 80 | 6.17 | 3.84 | 3.40 | 536 | 1.62 | 7159 | 2.1 | 1.13 |
| 90 | 6.42 | 3.99 | 3.14 | 627 | 1.68 | 9060 | 2.0 | 1.22 |
| 100 | 6.65 | 4.13 | 2.93 | 722 | 1.74 | 11186 | 1.8 | 1.31 |

## 4. Experimental Protocol

We present here an experimental protocol to realize a discrete time crystal based on a $^{39}$K BEC bouncing resonantly on an oscillating mirror. As an example, we focus on an $s = 30$ system and a 532 nm Gaussian potential light-sheet mirror with $\sigma_0 l_0 = 10$ μm, $V_0/V_{max}$=0.8, $V_{max} \approx 4.6 \times 10^3$. Other time crystals in the range $s \approx 20 - 100$ can be accessed using estimates of the optimal parameters given in Table 2.

(1) *Preparation of initial atom cloud.* We start with a BEC of $N \approx 5000$ $^{39}$K atoms in a large 1064 nm crossed optical dipole trap (CODT) [46] located at a drop height $h_0 l_0 \approx 145$ μm above the atom mirror. The longitudinal and transverse trap frequencies are adjusted to about 95 Hz to produce a spherical CODT, so that the standard deviation of the Gaussian atomic distribution $\sigma_z l_0 \approx 1.17$ μm matches the width of the Wannier wave-packet $|w_i(z,0)|^2$ at the classical turning point.

(2) *Release of BEC from optical dipole trap.* The longitudinal trapping potential is then switched off to release the BEC from the CODT to fall on to the repulsive light-sheet mirror in the presence of a vertical 1064 nm optical waveguide with confinement frequency 95 Hz.

(3) *Frequency and amplitude of oscillating mirror.* Next, the frequency of the oscillating mirror is tuned to the $s : 1$ resonance: $\omega = \frac{1}{2} s [g/(2h_0)]^{1/2}$ (e.g., $\omega/(2\pi t_0) = 2.8$ kHz for $s = 30$ and $h_0 l_0 \approx 145$ μm), and the amplitude of the mirror oscillation is set to $\lambda l_0/\omega^2 \approx 75$ nm (for a Gaussian potential mirror with $\sigma_0 l_0 = 10$ μm, $V_0/V_{max} \approx 0.8$) to create an energy gap $\Delta E/J \approx 10$. The light-sheet atom mirror is oscillated, e.g., by reflecting the beam off a piezo-driven optical mirror or possibly by modulating the intensity of the light-sheet beam.

(4) *Detection of time crystal.* The atom density is measured at fixed positions between the classical turning point and the atom mirror and at different moments in time out to typically $t/T = 2000$ mirror oscillations (or about 0.6 s). Two kinds of measurements are performed: (i) With the particle interaction set to zero ($g_{1D}N$=0), so that all atoms will have tunnelled out of the initially populated wave-packet into neighbouring wave-packets at $t/T = 2000$ to form a bunch of spatially separated wave-packets that spread with evolving time, as indicated in Figs. 3 and 4 of Ref. [6]. (ii) With the particle interaction sufficiently large to break the time-translation symmetry (e.g., $g_{1D}N=-0.2$), so that atoms will not tunnel from the initially populated wave-packet and the system evolves with period $sT$ and without decay out to at least $t/T = 2000$, as indicated in Figs. 3 and 4 of Ref. [6]. The first measurement (i) is needed in order to demonstrate that without a sufficiently strong particle interaction there is no ($sT$)-periodic time evolution that breaks the time translation symmetry.

## 5. Experimental Feasibility

Previous experiments have been successfully performed on a bouncing BEC dropped from heights $150 - 300$ μm on to a static 532 nm light-sheet atom mirror [38], similar to the atom mirror proposed here, and the phase coherence of the BEC was found to be preserved following multiple bounces. Similar results have been found for a thermal atom cloud bouncing from an evanescent-wave light mirror [47].

### 5.1 Lifetime of bouncing BEC

The lifetime of a bouncing BEC may be limited by losses due to three-body recombination collisions, two-body collisions with background atoms and molecules, photon scattering caused by stray near-resonant light, and atoms missing the mirror due to spreading of the wave-packet in the transverse directions or walk-off.

The lifetime due to three-body recombination is given by $\tau_{3b} = [\langle n^2 \rangle K_3]^{-1}$, where $K_3 = 1.3$ (5)$\times 10^{-29}$ cm$^6$s$^{-1}$ for $^{39}$K $|1, +1\rangle$ atoms near the zero-crossing point [39]; for example, for $\sigma_z l_0 = \sigma_\perp l_0 = 1.17$ μm (at the classical turning point) and $N=5000$ atoms, we obtain $\langle n^2 \rangle \approx 8 \times 10^{27}$ cm$^{-6}$ and hence $\tau_{3b} \approx 10$ (4) s. Losses due to collisions with background atoms can be kept to a negligible level by maintaining a high-quality vacuum ($< 10^{-11}$ millibar) and losses due to photon scattering should be negligible for light from the far-detuned 532 nm light-sheet mirror. Spreading of the atom cloud along the transverse directions will be suppressed due to the presence of the vertical optical waveguide. For atoms bouncing resonantly on an oscillating atom mirror, there will be no spreading in the longitudinal direction because each bounce from the oscillating mirror refocuses the atoms back [35]. In the calculations presented here, we assume about 50 bounces (corresponding to 1500 mirror oscillations for $s=30$ and lasting about 0.6 s) can be achieved without significant atom losses for drop heights of around 200 μm.

### 5.2 Spatial resolution

For $s = 30$, $\omega = 4.45$ and $\omega_z/(2\pi t_0) = \omega_\perp/(2\pi t_0) \approx 95$ Hz, the drop height $hl_0 \approx 145$ μm is much larger than the longitudinal extension of the atom cloud ($2\sigma_z l_0 \approx 2 - 10$ μm between the turning point and the atom mirror) and thus it should be possible to probe the position of the atom cloud with reasonable spatial resolution. Higher resolution could be achieved, if required, by choosing a higher $s$ resonance, for which the drop height (in SI units) scales as $s^{4/3}$ (Table 2).

*5.3 Amplitude of mirror oscillation*

For a soft Gaussian potential mirror with $\sigma_0 l_0 = 10$ μm, $V_0/V_{max} = 0.8$, $\omega = 4.45$, the optimal oscillation amplitude is $\lambda = 2.3$, or $\lambda l_0/\omega^2 = 74$ nm at a mirror oscillation frequency of 2.8 kHz for $^{39}$K in the laboratory frame. Still larger mirror oscillation amplitudes could be used, if required, by operating closer to the peak in Fig. 7(a) provided the corresponding classical resonance islands remain stable. Our estimates show that perturbations due to mechanical vibrations transmitted by a typical optical table are negligible when we want to control the motion of the atom mirror to a few nanometers at oscillation frequencies of a few kilohertz.

*5.4 Stray magnetic fields*

In the present experiment, the *s*-wave scattering length $a_s$ needs to be adjusted precisely to small negative values (e.g., $a_s l_0 = -1.6 a_0$ for $g_{1D}N = -0.2$ with $N = 5000$) for the time crystal phase or to zero for the non-time crystal phase. For the broad $^{39}$K $|1, +1\rangle$ Feshbach resonance, tuning the scattering length to a precision of $\pm 0.1 a_0$ requires the magnetic field to be stable to $\pm 0.18$ G which is much larger than the stray AC and DC magnetic fields achievable in the laboratory.

## 6. Discussion and Conclusions

We have investigated the range of sizes *s* of discrete time crystals that can be created for a BEC of attractively interacting atoms bouncing resonantly on an oscillating mirror. We have also considered the effects of having a realistic soft Gaussian potential mirror, such as that produced by a repulsive light-sheet, and suitable atomic systems for performing a time crystal experiment.

We find that for reflection from a soft Gaussian potential mirror the optimal amplitude of the mirror oscillations ($\sim 75$ nm for $^{39}$K atoms with $s = 30$, $\omega/(2\pi t_0) = 2.8$ kHz) is about an order of magnitude larger, and hence more experimentally accessible, than the optimal oscillation amplitude for a hard-wall mirror. For reflection from a Gaussian potential mirror the trajectories of the atoms are smooth at the position of reflection, so that smaller amplitude harmonics are created and, as a result, larger mirror oscillation amplitudes are needed to create a sufficiently large band gap ($\Delta E/J \approx 10$) and stable resonance islands, compared with a hard-wall potential mirror.

We find that using a $^{39}$K BEC and realistic experimental parameters, it should be possible to create discrete time crystals with sizes in the range $s \approx 20 - 100$. For $s < 20$, the drop height starts to become small ($< 80$ μm for the parameters considered), which makes it difficult to probe the atom density at different fixed positions with high spatial resolution, while for $s > 100$, the optimal mirror oscillation amplitude becomes small and the tunnelling times in the absence of interactions start to become long ($> 1.3$ s), which increases the time needed to perform an experiment compared with the lifetime of a bouncing BEC.

The robustness and stability of these time crystals against small perturbations are summarised in the form of a phase diagram of the detuning parameter $\Delta h_0$ against interaction strength $g_{1D}N$ in an accompanying paper in this issue [7].

Time crystals involving ultracold atoms bouncing resonantly on an atom mirror provide a platform for investigating a broad range of non-trivial condensed matter phenomena in the time domain. These include Mott insulator-like phases in the time-domain [26]; Anderson localization [6, 26] and many-body localization [27] due to temporal disorder; dynamical quantum phase transitions in time crystals [6, 28]; many-body systems with exotic long-range interactions [29]; time quasi-crystals – which are ordered but not periodic in time [29, 30]; and topological time crystals [31].

## Acknowledgements

We thank Zoran Hadzibabic for suggesting the use of $^{39}K$ atoms and Shannon Whitlock for fruitful discussions. Support of the Australian Research Council (DP190100815) and the National Science Centre Poland via Projects No. 2016/20/W/ST4/00314 and 2019/32/T/ST2/00413 (K.G.) and 2018/31/B/ST2/00349 (K.S.) is gratefully acknowledged. K.G. acknowledges the support of the Foundation for Polish Science (FNP).